\documentclass[runningheads]{llncs}
\pdfoutput=1
\usepackage[T1]{fontenc}
\usepackage{amsmath,amssymb,amsfonts}
\usepackage{cite}
\usepackage{algorithmic}
\usepackage{graphicx}
\usepackage{bm}
\usepackage{multirow}
\usepackage{booktabs}
\usepackage[ruled,linesnumbered]{algorithm2e}
\usepackage{setspace}
\usepackage{marvosym}
\usepackage{hyperref} 
\hypersetup{hidelinks} 
\usepackage{color}

\usepackage{placeins} 
\def\BibTeX{{\rm B\kern-.05em{\sc i\kern-.025em b}\kern-.08em
		T\kern-.1667em\lower.7ex\hbox{E}\kern-.125emX}}
\makeatletter
\def\thanks#1{\protected@xdef\@thanks{\@thanks
		\protect\footnotetext{#1}}}
\makeatother

\begin{document}
	\title{Debias the Black-box: A Fair Ranking Framework via Knowledge Distillation
	\thanks{Co-corresponding authors: Jianzong Wang (jzwang@188.com) and Yaodong Yang (yaodong.yang@pku.edu.cn). }
	}
	\author{Zhitao Zhu\inst{1,2} \and Shijing Si\inst{1,3} \and Jianzong Wang\inst{1}\inst{(}\textsuperscript{\Letter}\inst{)} \and Yaodong Yang\inst{4}\inst{(}\textsuperscript{\Letter}\inst{)} \and Jing Xiao\inst{1}}
	\authorrunning{Zhitao Zhu et al.}
	\institute{Ping An Technology (Shenzhen) Co., Ltd., Shenzhen, China \\
		\email{andyzzt@mail.ustc.edu.cn, shijing.si@outlook.com, jzwang@188.com, xiaojing661@pingan.com.cn}
		\and  IAT,	University of Science and Technology of China, Hefei, China\\
		 \and 
		 School of Economics and Finance, Shanghai International Studies University, Shanghai, China \\
		 \and
		 Institute for AI, Peking University, Beijing, China \\
		 \email{ yaodong.yang@pku.edu.cn}}
	\maketitle              
	\begin{abstract}
		Deep neural networks can capture the intricate interaction history information between queries and documents,  because of their many complicated nonlinear units, allowing them to provide correct search recommendations. However, service providers frequently face more complex obstacles in real-world circumstances, such as deployment cost constraints and fairness requirements. Knowledge distillation, which transfers the knowledge of a well-trained complex model (teacher) to a simple model (student), has been proposed to alleviate the former concern, but the best current distillation methods focus only on how to make the student model imitate the predictions of the teacher model. To better facilitate the application of deep models, we propose a fair information retrieval framework based on knowledge distillation. This framework can improve the exposure based fairness of models while considerably decreasing model size. Our extensive experiments on three huge datasets show that our proposed framework can reduce the model size to a minimum of 1\% of its original size while maintaining its black-box state. It also improves fairness performance by 15\%\textasciitilde46\% while keeping a high level of recommendation effectiveness.
		
	\end{abstract}
	
	
	\section{Introduction}\label{sec:intro}
	Information Retrieval (IR) systems are nowadays one of the most pervasive techniques in a variety of industries. 
	The sophisticated architectures and growing data scale of application scenarios cause the size of models to increase rapidly.
	 Large models tend to capture complicated interactions between queries and documents, yielding increased performance at the expense of increased computational time and memory phase.
	
\begin{figure}[!ht]
	\centering
	\includegraphics[width=0.9\linewidth]{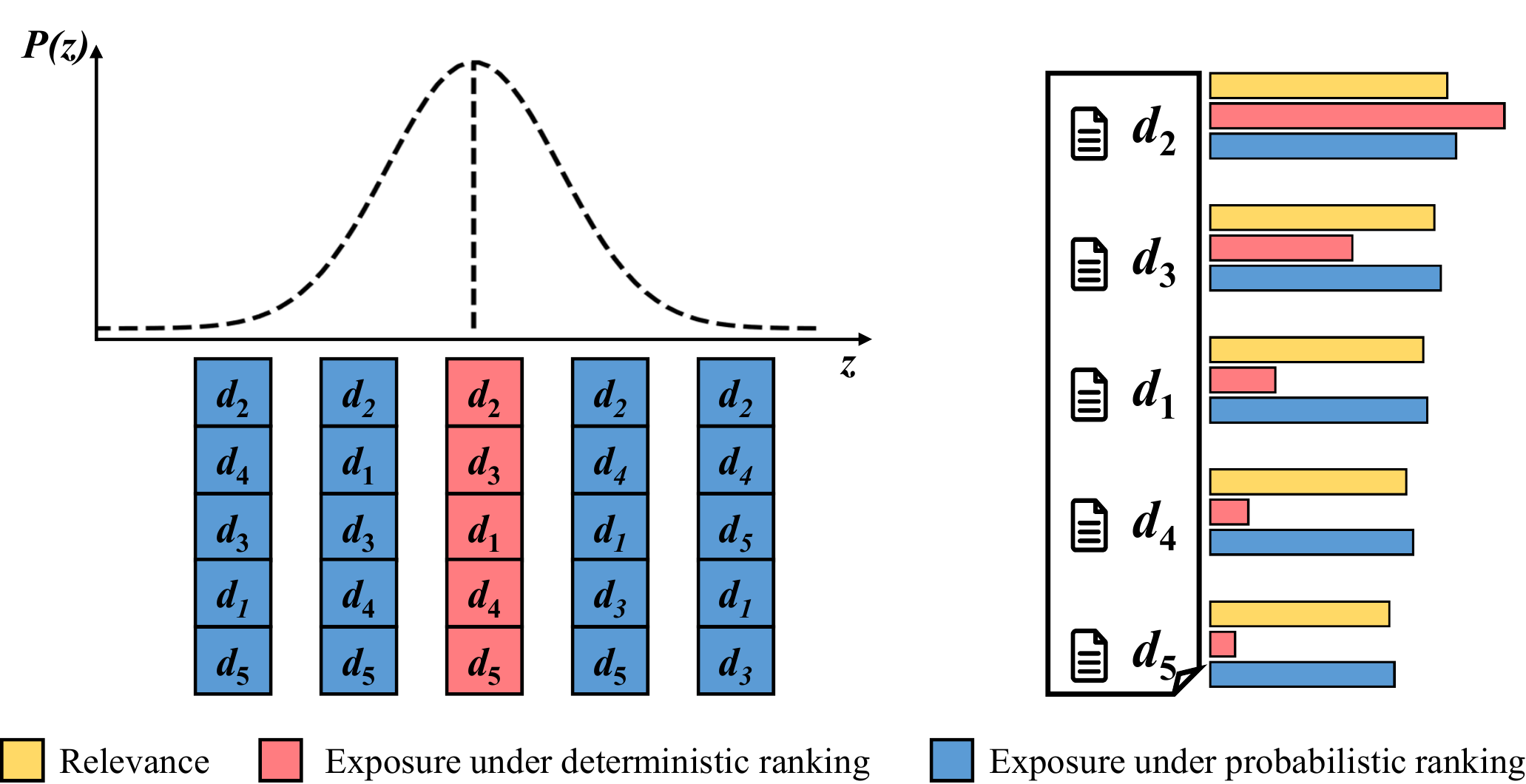}
	\caption{A simple ranking method that relies only on ordering relevance scores makes the ratio between exposure and relevance disproportionate, magnifying subtle gaps in relevance. In contrast, a probabilistic ranking model maintains a positive relationship between exposure and relevance. $z$ denotes possible permutations.}
	\label{exposure}
\end{figure}

	To tackle the difficulty of applying such large models to web-scale and real-time platforms, a few recent works \cite{Tang2018RankingDL,Lee2019CollaborativeDF,Kweon2021BidirectionalDF,Kang2020DERRDAK} have applied Knowledge Distillation (KD) \cite{DBLP:journals/corr/HintonVD15} to IR. Most ranking distillation approaches, however, focus on the balance of prediction performance and computing speed. Little attention has been paid to the fairness of ranking models during the distillation process \cite{18SinghJfairnessexposure}. 
	Ranking systems employ deterministic models to assign an individual score to each item, and then sort items in descending order of their assigned scores to obtain rankings. That calculation pattern is succinct and intuitive,  yet, leads to unfairness in the distribution of exposure. Exposure represents the expected number of people who will check an item. User behavior is affected by position bias:  they are less likely to check items at lower positions. As illustrated in Figure \ref{exposure}, this results in the allocation of user attention being disproportionate to the rankings on the recommendation list.
\vspace{-0.4cm}
	A small difference in relevance can make a world of difference in exposure as a result of the winner-take-all  \cite{18SinghJfairnessexposure} allocation of exposure. Under this circumstance, the unfairness between candidates is significantly amplified. Furthermore, while the combination of deep learning with ranking models provides a fresh leap forward, its lack of interpretability and increased reliance on data may introduce endogenous biases. The trade-off between model performance and fairness has become a huge issue for ranking models.
	
	In this research, we propose a general Fairness-aware Ranking Distillation framework (FRD) for improving the fairness of top-$K$ ranking models. FRD eliminates the average disparity of exposures documents received, and it is a direct approach irrelevant to specific protected attributes. Our framework takes advantage of ranking distillation, so it can build on top of large well-trained ranking models without extra retraining cost.
	The main contributions of this work can be summarized as follows:
	\begin{itemize}
		\item We adopt a ranking distillation framework with sub-regional treatment that retains the top rankings while penalizing items ranked lower by teachers. Since only the soft label results of the teacher model are used, our framework can be effectively applied to the black-box models.

		\item We identify the great potential of ranking distillation and propose that implementing a personalized fairness correction in the process of ranking distillation can avoid the vast additional computational costs necessary to achieve fairness purposes directly on complex models.

		\item By applying the latest optimization method of the PL ranking model, we have further reduced the computational cost of fairness correction significantly.
	\end{itemize}
	
	\section{Related Work}
	\label{sec:related}	
	\subsection{Knowledge Distillation}
	The initial KD method transfers knowledge through the softmax output of a teacher model. Some subsequent work has extended the scope of statistics used for matching, such as  intermediate feature responses \cite{DBLP:conf/nips/DentonZBLF14}, gradient \cite{DBLP:conf/icml/SrinivasF18}, and distribution \cite{DBLP:journals/corr/HuangW17a}, etc. 
	Other work, on the other hand, opted for a more subtle structural design. For example, DE-RRD \cite{Kang2020DERRDAK} uses distillation experts to learn the middle layer representation space mapping function of Teacher while additionally using the rank order given by Teacher for rank matching; DCD \cite{Lee2021DualCS} assigns more training resource to instances that were not correctly predicted by Student; MiniLM \cite{DBLP:conf/nips/WangW0B0020} proposes an assistant mechanism to distill only the last layer of the self-attentive matrix and the value-value matrix of the pre-trained model. But the potential of KD for other objectives remains largely unexplored.
	
\subsection{Fairness in rankings}
Despite the growing impact of online information systems on our society and economy, the fairness of rankings has been a relatively under-explored area.
In the existing work, some consider the equity of groups with respect to a set of categorical sensitive attributes (or features) to be ranked according to the principle of population parity, which can be further divided into group fairness \cite{DBLP:conf/nips/KusnerLRS17} and individual fairness \cite{DBLP:conf/iclr/MaityXYS21}. 
But there are other works such as \cite{18SinghJfairnessexposure,Diaz2020EvaluatingSR} that argue that the fairness of ranking systems corresponds to how they assign exposure based on the merits of individual items or item groups. 
These works specify and enforce fairness constraints that explicitly link relevance to exposure in expectation or amortized over a set of queries \cite{Singh2019PolicyLF}. 
	\section{Methodology}
	\label{sec:RankDistil}
	In this section, we illustrate FRD, our universal fairness-aware ranking distillation framework, which is made up of ranking distillation and a fairness penalty. We elaborate on these two components in this section and present the algorithm for FRD. 
		\begin{figure*}[!ht]
		\centering
		\centerline{\includegraphics[width=1\linewidth]{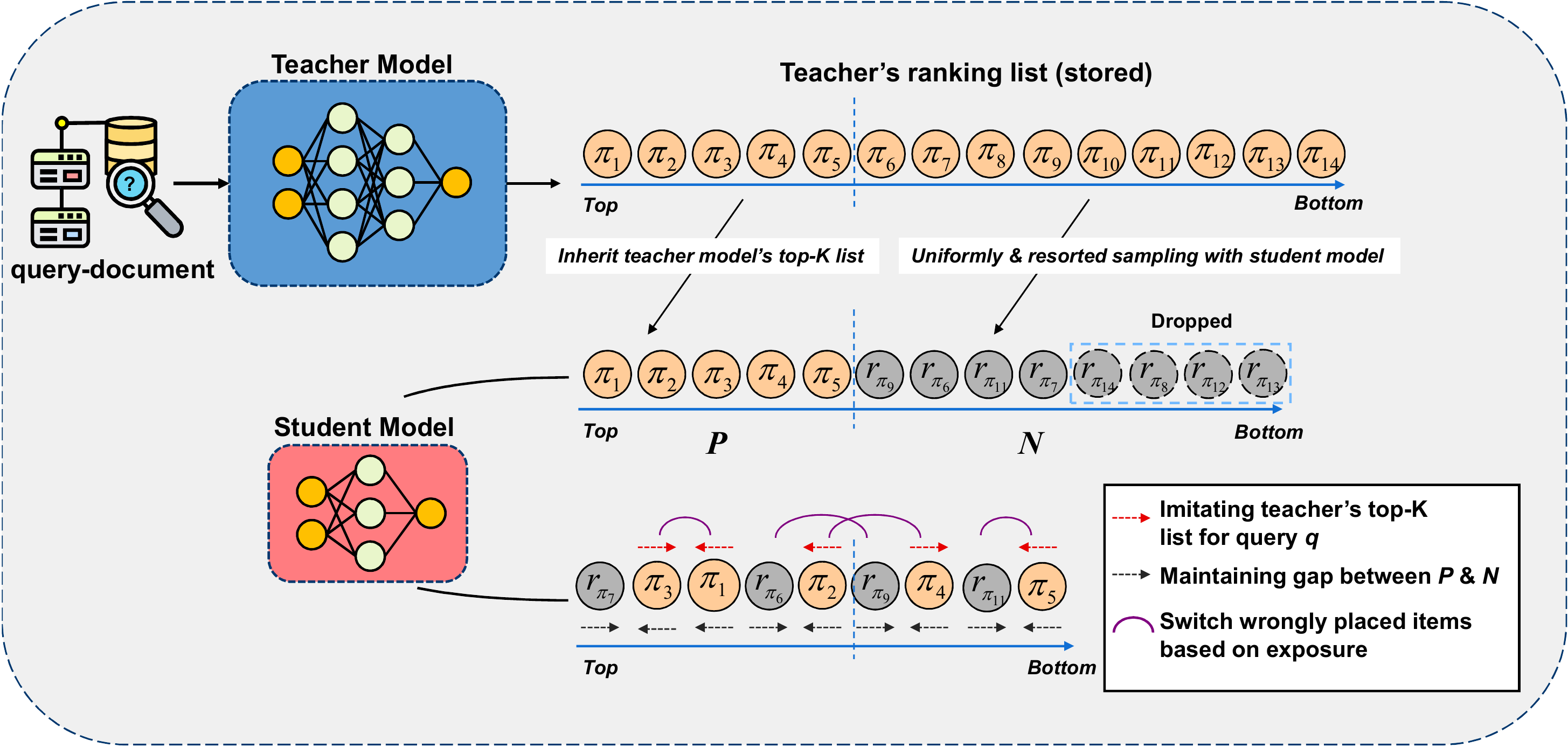}}
		\caption{Illustration of FRD framework. Student studies to rearrange the items into two sets to (a). maintain the same top-$K$ list as the teacher; (b). penalize negative set items that are ranked too high; (c). ensure that items expected to get more exposure are placed higher up the list by pairwise comparison}
	\label{framework}
\end{figure*}
	 	\vspace{-1cm}
\subsection{Ranking Distillation}
	Ranking distillation attempts to mimic the teacher ranking model by transferring its knowledge to the student. 
	For we only care about the top performers, considering the scores of all query-document pairs is undoubtedly inefficient and wasteful. A straightforward approach is to use only the information of  top-$K$ items and discard items ranked low by the teacher like \cite{Tang2018RankingDL}. But this approach leads to performance degradation because some useful information is discarded. 
	It is undesirable to discard directly or incorporate wholesale into the training for candidate items that are not on the teacher model recommendation list.

	1). For items recommended by Teacher, student network learns to match all their orders to get fine-grained scoring results.
	
	2). Candidates not recommended by the teacher model are also internally differentiated: items with very low scores do not help maintain a distinct score boundary, and the model should be allowed to learn more information from candidates close to the boundary.
	
	FRD treats Teacher's top-$K$ candidates as positive set $P$ to help the student imitate top performers' order and samples out negative set $N$ to keep less relevant documents at lower end. Inspired by \cite{reddi2021rankdistil}, two categorical ranking losses $\mathcal{L}_{P}(t,s,P)\text{ and }\mathcal{L}_N(s,P,N)$ are employed to allocate computing resources rationally according to importance, where $t,\ s \in \mathbb{R}^A$ represent the score list generated by Teacher and Student respectively. For the top-K list containing more refined sorting information of $f^T$, we construct a softmax based cross-entropy loss to minimize the difference between student's and teacher's top-p lists (positive set $P$). It is worth mentioning that here we relax $K$ to $p$ in Algorithm \ref{al:FRD} to give more items recommendation opportunities.  
	\begin{equation}
		\mathcal{L}_P(t,s,P) = \sum_{i \in P}\left\{-\frac{e^{t_i}}{\delta^t_P} \cdot \log \frac{e^{s_{i}}}{\delta^s_P}\right\} \text{ where } \delta_P^t=\sum_{j \in P} e^{t_j}
	\end{equation}
	
	Later, for the large negative sample set, the binary or pairwise comparison loss functions like Equation \eqref{eq:distill} with less complexity can be employed since we only need to discriminate the rest items without considering the precise ranking. A negative sampling strategy is adopted to further reduce the number of negative candidates. A compressed set $C$ is first sampled out with a categorical distribution $Q$, and then score with the student model to pick the top $n$ documents to form the negative set $N$. This process brings a huge reduction in the calculation amount but no loss of effect. 
	
	\begin{equation}
		\label{eq:distill}
		\begin{aligned}
			\mathcal{L}_{N_b}(s,N)&=\sum_{i\in N} \log (1+e^{s_i})\text{\ \ \ \ \ \ \ \ \ \ \ \ \ \  (Binary)}\\
			\mathcal{L}_{N_p}(s,P,N)&=\sum_{i\in N} \sum_{j\in P} \log (1+e^{s_i-s_j})\text{\ \ \ (Pairwise)}
		\end{aligned}
	\end{equation}
	This categorical distillation strategy helps us concentrate finite computational power on the most useful retrieved list displayed to users while making good use of the remaining items.
\vspace{-0.2cm}
	\subsection{Exposure-based Fair Ranking}
	It is widely accepted that an item's position in the rankings has a critical impact on its exposure and financial success. Yet, surprisingly, the algorithms used to learn these rankings are often blind to their impact on items.
	The key goal of our amelioration is to promote exposure-based fairness during distillation. Based on the exposure estimating approach proposed below, we can explicitly specify how exposure is allocated such as  making exposure proportional to relevance.
	
	In recent years, the popularity of the Plackett-Luce (PL) ranking model \cite{luce2012individual} has increased. 
	It is more efficient to use the PL ranking model to model the probabilistic distribution over rankings directly. Especially for fair distributions of attention exposure, multiple lines of previous works have found that PL ranking models may transfer the deterministic recommendation pattern into reflecting the relevance by the probability of occurrence \cite{18SinghJfairnessexposure,Diaz2020EvaluatingSR,Singh2019PolicyLF}, which gives an approximately equal possibility of being the top-item to candidates with slightly lower relevance indicators.
	
In the meantime, we follow the definition of exposure-based fairness: the items with higher ranking should achieve more exposure in practical environment. Thus we can constraint the difference between existing arrangement and counterfactual arrangement by minimize their exchanged exposure difference.
For brevity we denote $\varepsilon_d=\varepsilon(q,d)$ as the exposure an item $d$ receives under a probabilistic model $\pi$:
\vspace{-0.2cm}
\begin{equation}
	\varepsilon _d =\sum_{z\in\pi}\pi(z)\sum_{k=1}^K\theta_k\mathbb{I}[z_k=d] 
\end{equation}
where $z$ denotes a possible permutation and rank weight $\theta_k$ indicates the probability that a user examines the item at rank $k$ (e.g. $\frac{1}{k}$). 
For each query, this statistic calculates the averaged disparity between every two items and takes the average disparity across all item pairs:
\label{sec:fair}
\vspace{-0.1cm}
\begin{equation}\label{eq:disp}
	\mathcal{L}_{Fair}(s,P)=\cfrac{1}{|P|(|P|-1)}\sum_{d\in P}\sum_{d'\in P}(\varepsilon_{d'}\mathcal{R}_d-\varepsilon_d\mathcal{R}_{d'})^2
\end{equation}
\vspace{-0.1cm}
Here the relevance of a document $\mathcal{R}_d$ is set as a transformation of its label and the exposures are estimated by $d$'s total possibly gained rank weight \cite{Oosterhuis2021ComputationallyEO} calculated through a continued product of following softmax principle:
\vspace{-0.1cm}
\begin{equation}
	\pi(d|z_{1:k-1,A})=\frac{e^{s(d)}\cdot\mathbb{I}[d\notin z_{1:k-1}]}{\sum_{d'\in A \backslash z_{1:k-1} }e^{s(d')}}    
\end{equation}
where $s(d)$ denotes the score that the student model assigns to document $d$.
	
Calculating the gradient of a PL ranking model, on the other hand, necessitates iterating through every conceivable ranking that the model may generate, i.e, every possible permutation. In reality, this computational impossibility is overcome by estimating the gradient using model rankings as samples \cite{DBLP:conf/wsdm/OosterhuisR21}.
Employing the estimator to the gradient of PL model proposed in  \cite{Oosterhuis2021ComputationallyEO}, we can optimize the exposure-based unfairness metrics simultaneously during knowledge distillation. 
This Metric \eqref{eq:disp} is designed to measure the difference in reward if the exposures of two items were swapped. Substituting the derivative of this metric with respect to $\varepsilon_d$, we can employ the below-mentioned estimator \cite{Oosterhuis2021ComputationallyEO} to estimate its gradient and then update the PL model by back propagation, $m$ denotes the ranking merit and an entire ranking is sampled with Gumble Softmax \cite{gumbel1954statistical} trick rather than calculating any of the actual probabilities \cite{DBLP:conf/wsdm/BruchHBN20}.
 \begin{equation}
	 	\begin{aligned}
		 		\cfrac{\delta }{\delta m} \mathcal{M}etric(q)= &\sum_{d\in P}\Big[ \cfrac{\delta }{\delta m }m(d)\Big]\mathbb{E}_z[(\sum_{k=\text{rank}(d,z)+1}^K\theta_k[ \cfrac{\delta \mathcal{M}etric(q)}{\delta \varepsilon_{z_k}}]\\
		 		&+\sum_{k=1}^{rank(d,z)}\pi(d|z_{1:k-1)}(\theta_k[\cfrac{\delta \mathcal{M}etric(q)}{\delta \varepsilon_{z_k}}]
		 	\end{aligned}
	 \end{equation}
\vspace{-0.8cm}
\begin{algorithm}[ht]
	\setstretch{1}
	\caption{FRD}
	\label{al:FRD}
	\begin{algorithmic}[1]
		\STATE Student model $f^S$ and well-trained Teacher model $f^T$. \\
		Hyper-parameters: Positive set size $p\in[K,a]$, Compressed set size $c\leq a-p$, Negative set size $n\leq c $, Learning rate $\eta$, Epochs $R$.\\
		\FOR{each $r \in [0,R-1]$}
		\STATE Uniformly randomly select an query $x$, denote its  candidate set of $a$ documents as $A$ with stored teacher scores $t$\
		\STATE \texttt{/* Indices of vector's largest $p$ elements */}\\
		Compute positive set $P=Top_p(t)$\
		\STATE Randomly sample the compressed set $C$ of $c$ items with $Q\left(A-P\right)$\
		\STATE Compute negative set $N=Top_n\left(f^S(C)\right)$\
		\STATE Compute student scores $s=f^S\left(P\bigcup N\right)$
		\STATE Update $f^S$ by optimizing the total loss in Equation \eqref{eq:total.loss} \
		\ENDFOR
		\label{code:recentEnd}
	\end{algorithmic}
\end{algorithm}

In total, the loss of FRD is
\begin{equation}\label{eq:total.loss}
	\mathcal{L}=\mathcal{L}_{P}(t,s,P)+\mathcal{L}_N(s,P,N)+\lambda \mathcal{L}_{Fair}(s,P)
\end{equation}
where $\lambda$ is the tuning parameter specific to the dataset.
This framework is model-agnostic as it only requires the teacher network's scoring data, making it easy to distill black box models. It is also able to debias ranking models against many attributes as it does not require any side information.
\vspace{-0.2cm}
\section{Experiments}
\label{sec:result}

In this section, we show empirical performance for FRD. We compare our method with the other baselines on two aspects: 1). ranking performance through imitating teacher; 2). unfairness correction level during distillation.

\subsection{Datasets and Training Configurations}
\noindent{\textbf{Datasets}.} We conduct our experiments on three public datasets: MSLR-WEB30K (fold 1) \cite{DBLP:journals/corr/QinL13}, Yahoo LETOR \cite{yahooletorChapelleC11} and Istella LETOR full dataset \cite{DBLP:conf/sigir/LuccheseNOPTV16}, which are all benchmark datasets for large-scale experiments on the efficiency and scalability of LTR solutions. The basic statistics of these datasets can be found in \cite{reddi2021rankdistil}.

\noindent{\textbf{Teacher/Student Models}.} To better control the ratio of distillation, we employ the Multi-Layered Perceptron (MLP) as the architecture for both the teacher and student networks.
Specifically, the teacher model is a 3 layer FC-BN-ReLU model with hidden units of sizes 1024, 512, 256 for all three datasets and the student models are chosen by varying the proportion of parameter counts of the teacher and student models.

\noindent{\textbf{Evaluation Protocols.}}  Following \cite{Tang2018RankingDL,reddi2021rankdistil}, we measure the ranking performance of student models in terms of the normalized discounted cumulative gain (NDCG) at top 5 \& 10 , which is commonly used for ranking tasks. 
For the fairness of a ranking model, we utilize the averaged squared disparity in Equation \eqref{eq:disp} over all queries as the disparity metric. 

\subsection{Ranking and Fairness Performance}
The distillation and fairness performances of teacher and student models on three datasets are presented in Table \ref{tab:distill_performance}. The student model is a two-layered FC-BN-ReLU model of hidden units of sizes 28, 28. We utilize two kinds of distillation loss, the pairwise and binary loss, shown in methods ending with `P' and `B', respectively.

FRD consistently outperforms RankDistil and the teacher model in fairness. The averaged disparity of FRD methods decreases by 15\%\textasciitilde46\%, compared with the teacher model. The student models of RankDistil also exhibit a slight improvement in fairness, which is caused by the reduced size of the model. 
Note that we attribute to the differences unfairness values due to the different composition of the datasets, for example, each query in the Istella test set corresponds to an average of 319 documents, while for the Yahoo dataset, the number is 24. Deeper reasons remain to be explored, but for now, this fairness metric is not applicable to cross-dataset comparisons.
\vspace{-0.2cm}
\begin{table}[!htbp]
	\centering
	\fontsize{9}{15}
	\caption{Performance of various distillation methods on three datasets. Methods ending with `B' indicate that binary loss is used for distillation, and `P' denotes pairwise distillation loss. `Disp.' denotes the averaged disparity over all queries.}\label{tab:distill_performance}
	\renewcommand\arraystretch{1}  
	\setlength{\tabcolsep}{0.3mm}{
		\centerline{
			\begin{tabular}{lcccccccccccc}
				\toprule[1.25pt]
				\multirow{2}{*}{Method/Data} &  & \multicolumn{3}{c}{MSLR Web30K} &  & \multicolumn{3}{c}{Yahoo LETOR} &  & \multicolumn{3}{c}{Istella} \\ \cline{3-5} \cline{7-9} \cline{11-13} 
				&  & N$_5\uparrow$     & N$_{10}\uparrow$    & Disp.$\downarrow$    &  & N$_5\uparrow$     & N$_{10}\uparrow$    & Disp.$\downarrow$    &  & N$_5\uparrow$     & N$_{10}\uparrow$    & Disp.$\downarrow$   \\ \toprule[1.25pt]
				Teacher                      &  & 0.467     & 0.488     & 0.080        &  & 0.722     & 0.764     &  0.647       &  & 0.628   & 0.684    &  0.013      \\ \hline
				Ranking  Distillation \cite{Tang2018RankingDL} & & 0.428 & 0.451&0.080&&0.684&0.731&0.641&&\textbf{0.490}&0.525&0.013\\
				RankDistil-B \cite{reddi2021rankdistil}               &  & 0.423     & 0.457     &  0.075       &  & 0.683     & 0.725     &  0.621       &  & \textbf{0.490}   & 0.525    & 0.011       \\
				RankDistil-P \cite{reddi2021rankdistil}              &  & \textbf{0.457 }    & \textbf{0.479}     & 0.076        &  & \textbf{0.710 }    & \underline{0.750}     &  0.625       &  & 0.481   & \textbf{0.528}    &  0.012     \\ 
				FRD-B(\textit{ours})               &  & 0.420     & 0.452     & 0.063        &  & 0.678     & 0.721     & 0.551        &  & 0.483   & 0.520    &  0.008      \\
				FRD-P(\textit{ours})                &  & \underline{0.455}     & \underline{0.478}     & \textbf{0.059}        &  & \underline{0.706}     & \textbf{0.751}     & \textbf{0.550}        &  & \underline{0.487}   & \underline{0.526}    &  \textbf{0.007}      \\ \bottomrule[1.25pt]
			\end{tabular}
	}}
\end{table}
\vspace{-0.2cm}
With regards to the ranking performance, student models trained with pairwise loss significantly outperform binary loss. Both RankDistil and FRD with pairwise loss yield similar performances to the teacher model on MSLR Web30K and Yahoo LETOR datasets. On Istella dataset, student models perform significantly worse than the teacher model due to their lack of  capacity  to capture enough patterns from the huge dataset.
\vspace{-0.2cm}
\subsection{Performance of Student Models versus Size}
Table \ref{tab:performance_vs_size} displays the performance of pairwise distillation methods on two datasets while varying the size of student models from 50\% to 1\% of Teacher. We vary the size of student models by adjusting the number of hidden neuron units. Fewer parameters means faster inference.

\begin{table}[h!]
	\centering
	\caption{Performance of distillation methods while varying the size of student models. The `NDCG' is evaluated on top-5 documents, and `Disp.' is the averaged disparity over all queries.}\label{tab:performance_vs_size}
	\renewcommand\arraystretch{1}  
	\setlength{\tabcolsep}{1.2mm}{
		\begin{tabular}{llcccccc}
			\toprule[1.25pt]
			\multirow{2}{*}{Models} & \multirow{2}{*}{Size} &  & \multicolumn{2}{l}{MSLR Web30K} &  & \multicolumn{2}{l}{Yahoo LETOR} \\ \cline{4-5} \cline{7-8} 
			&                       &  & N$_5\uparrow$          & Disp. $\downarrow$          &  & N$_5\uparrow$          & Disp. $\downarrow$           \\ \toprule[1.25pt]
			Teacher                 & 100\%                 &  & 0.467          & 0.080          &  & 0.722          & 0.647          \\ \hline
			RankDistil-P           & 50\%                  &  & 0.435          & 0.079          &  & 0.721          & 0.635          \\
			FRD-P          & 50\%                  &  & 0.433          & 0.069          &  & 0.720          & 0.578          \\
			RankDistil-P           & 10\%                  &  & 0.431          & 0.078          &  & 0.721          & 0.612          \\
			FRD-P          & 10\%                  &  & 0.428          & 0.064          &  & 0.719          & 0.563          \\
			RankDistil-P           & 1\%                   &  & 0.422          & 0.074          &  & 0.715          & 0.589          \\
			FRD-P          & 1\%                   &  & 0.421          & \textbf{0.059}          &  & 0.714          & \textbf{0.550}          \\ \bottomrule[1.25pt]
	\end{tabular}}
\end{table}

Both RankDistil and FRD methods with pairwise loss are robust to the size of student models, as the NDCG$_5$ scores of the 1\% student models are comparable to the teachers. FRD can reduce the average squared disparity significantly faster than RankDistil, which indicates that the fairness penalty plays an essential role in alleviating the disparity in ranking. The distillation performance of FRD has some decrease compared to RankDistil, but not more than 0.06\%, which is negligible considering the improvement in fair performance.
\FloatBarrier
\vspace{-0.2cm}
\section{Conclusion}
\label{sec:conclusion}
\vspace{-0.2cm}
In this paper, we propose FRD, a fairness-aware ranking distillation framework that leverages the potential of knowledge distillation. It can inherit the excellent ranking retrieval capabilities of the teacher model in the black-box state, while reducing the model size to a minimum of one percent. We pioneer the use of KD in achieving the exposure fairness.
 Our experiments show that the FRD with a bias correction strategy can achieve a significant reduction in model size and a large improvement in the reasonable distribution of exposure.
\vspace{-0.2cm}
\subsubsection{Acknowledgements}
Zhitao Zhu and Shijing Si contributed equally to this work. Co-corresponding authors: Jianzong Wang and Yaodong Yang . This work is supported by the Key Research and Development Program of Guangdong Province under grant No.2021B0101400003. 
\vspace{-0.2cm}
\footnotesize
%
%
\bibliographystyle{splncs04}
\bibliography{refs}

\begin{thebibliography}{10}
\providecommand{\url}[1]{\texttt{#1}}
\providecommand{\urlprefix}{URL }
\providecommand{\doi}[1]{https://doi.org/#1}

\bibitem{DBLP:conf/wsdm/BruchHBN20}
Bruch, S., Han, S., Bendersky, M., Najork, M.: A stochastic treatment of
  learning to rank scoring functions. In: WSDM. pp. 61--69. {ACM} (2020)

\bibitem{yahooletorChapelleC11}
Chapelle, O., Chang, Y.: Yahoo! learning to rank challenge overview. In:
  Proceedings of the Yahoo! Learning to Rank Challenge, held at {ICML}. {JMLR}
  Proceedings, vol.~14, pp. 1--24. JMLR.org (2011)

\bibitem{DBLP:conf/nips/DentonZBLF14}
Denton, E.L., Zaremba, W., Bruna, J., LeCun, Y., Fergus, R.: Exploiting linear
  structure within convolutional networks for efficient evaluation. In: NeuIPS.
  pp. 1269--1277 (2014)

\bibitem{Diaz2020EvaluatingSR}
Diaz, F., Mitra, B., Ekstrand, M.D., Biega, A.J., Carterette, B.: Evaluating
  stochastic rankings with expected exposure. In: CIKM. pp. 275--284. {ACM}
  (2020)

\bibitem{gumbel1954statistical}
Gumbel, E.J.: Statistical theory of extreme values and some practical
  applications: a series of lectures, vol.~33. US Government Printing Office
  (1954)

\bibitem{DBLP:journals/corr/HintonVD15}
Hinton, G.E., Vinyals, O., Dean, J.: Distilling the knowledge in a neural
  network. In: NeuIPS. pp.~1--9. MIT Press (2015)

\bibitem{DBLP:journals/corr/HuangW17a}
Huang, Z., Wang, N.: Like what you like: Knowledge distill via neuron
  selectivity transfer. CoRR  \textbf{abs/1707.01219} (2017)

\bibitem{Kang2020DERRDAK}
Kang, S., Hwang, J., Kweon, W., Yu, H.: {DE-RRD:} {A} knowledge distillation
  framework for recommender system. In: CIKM. pp. 605--614. {ACM} (2020)

\bibitem{DBLP:conf/nips/KusnerLRS17}
Kusner, M.J., Loftus, J.R., Russell, C., Silva, R.: Counterfactual fairness.
  In: NeurIPS. pp. 4066--4076 (2017)

\bibitem{Kweon2021BidirectionalDF}
Kweon, W., Kang, S., Yu, H.: Bidirectional distillation for top-k recommender
  system. In: WWW. pp. 3861--3871. {ACM} / {IW3C2} (2021)

\bibitem{Lee2019CollaborativeDF}
Lee, J., Choi, M., Lee, J., Shim, H.: Collaborative distillation for top-n
  recommendation. In: ICDM. pp. 369--378. {IEEE} (2019)

\bibitem{Lee2021DualCS}
Lee, Y., Kim, K.E.: Dual correction strategy for ranking distillation in top-n
  recommender system. In: CIKM. pp. 3186--3190. ACM (2021)

\bibitem{DBLP:conf/sigir/LuccheseNOPTV16}
Lucchese, C., Nardini, F.M., Orlando, S., Perego, R., Tonellotto, N.,
  Venturini, R.: Exploiting {CPU} {SIMD} extensions to speed-up document
  scoring with tree ensembles. In: SIGIR. pp. 833--836. {ACM} (2016)

\bibitem{luce2012individual}
Luce, R.D.: Individual choice behavior: A theoretical analysis. Courier
  Corporation (2012)

\bibitem{DBLP:conf/iclr/MaityXYS21}
Maity, S., Xue, S., Yurochkin, M., Sun, Y.: Statistical inference for
  individual fairness. In: ICLR. pp. 1--19. OpenReview.net (2021)

\bibitem{Oosterhuis2021ComputationallyEO}
Oosterhuis, H.: Computationally efficient optimization of plackett-luce ranking
  models for relevance and fairness. In: SIGIR. pp. 1023--1032. {ACM} (2021)

\bibitem{DBLP:conf/wsdm/OosterhuisR21}
Oosterhuis, H., de~Rijke, M.: Unifying online and counterfactual learning to
  rank: {A} novel counterfactual estimator that effectively utilizes online
  interventions. In: WSDM. pp. 463--471. {ACM} (2021)

\bibitem{DBLP:journals/corr/QinL13}
Qin, T., Liu, T.: Introducing {LETOR} 4.0 datasets. CoRR
  \textbf{abs/1306.2597}, ~1--6 (2013)

\bibitem{reddi2021rankdistil}
Reddi, S., Pasumarthi, R.K., Menon, A., Rawat, A.S., Yu, F., Kim, S., Veit, A.,
  Kumar, S.: Rankdistil: Knowledge distillation for ranking. In: AISTATS. pp.
  2368--2376. PMLR, JMLR (2021)

\bibitem{18SinghJfairnessexposure}
Singh, A., Joachims, T.: Fairness of exposure in rankings. In: SIGKDD. pp.
  2219--2228. {ACM} (2018)

\bibitem{Singh2019PolicyLF}
Singh, A., Joachims, T.: Policy learning for fairness in ranking. In: NeurIPS.
  pp. 5427--5437. MIT Press (2019)

\bibitem{DBLP:conf/icml/SrinivasF18}
Srinivas, S., Fleuret, F.: Knowledge transfer with jacobian matching. In: ICML.
  Proceedings of Machine Learning Research, vol.~80, pp. 4730--4738. {PMLR}
  (2018)

\bibitem{Tang2018RankingDL}
Tang, J., Wang, K.: Ranking distillation: Learning compact ranking models with
  high performance for recommender system. In: SIGKDD. pp. 2289--2298. {ACM}
  (2018)

\bibitem{DBLP:conf/nips/WangW0B0020}
Wang, W., Wei, F., Dong, L., Bao, H., Yang, N., Zhou, M.: Minilm: Deep
  self-attention distillation for task-agnostic compression of pre-trained
  transformers. In: NeurIPS (2020)

\end{thebibliography}
\end{document}